\documentclass[conf]{new-aiaa}
\usepackage[utf8]{inputenc}

\usepackage{graphicx}
\usepackage{hyperref}
\usepackage{amsmath}
\usepackage[version=4]{mhchem}
\usepackage{siunitx}
\usepackage{longtable,tabularx}
\usepackage{tikz}
\newcommand{\bsu}{\boldsymbol{u}}
\newcommand{\bsq}{\boldsymbol{q}}
\newcommand{\bsf}{\boldsymbol{f}}
\newcommand{\bsx}{\boldsymbol{x}}
\newcommand{\bsU}{\boldsymbol{U}}

\DeclareMathOperator*{\argmax}{arg\max}

\usepackage{floatrow}
\usepackage{xcolor}
\definecolor{myOrange}{rgb}{0.8500,0.3250,0.0980}
\definecolor{myYellow}{rgb}{0.9290,0.6940,0.1250}
\definecolor{myGreen}{rgb}{0.4660,0.6740,0.1880}
\definecolor{myBlue}{rgb}{0,0.4470,0.7410}

\newcommand{\blackDashed}{\raisebox{2pt}{\tikz{\draw[black,dashed,line width=1.2pt](0,0) -- (5mm,0);}}}

\graphicspath{{Figures/}}

\begin{document}

\title{
Resolvent analysis of laminar and turbulent duct flows
}

\author{Barbara Lopez-Doriga\footnote{Graduate Research Assistant, MMAE Department - Illinois Institute of Technology, Chicago, IL 60616.} and Scott T. M. Dawson\footnote{Assistant Professor, MMAE Department - Illinois Institute of Technology, Chicago, IL 60616. AIAA Senior Member.}}
\affil{Illinois Institute of Technology, Chicago, IL 60616}
\author{Ricardo Vinuesa\footnote{Associate Professor, FLOW, KTH Engineering Mechanics, SE-100 44 Stockholm, Sweden.}}
\affil{FLOW, Engineering Mechanics, KTH Royal Institute of Technology,  Stockholm, Sweden.}
\graphicspath{{Figures/}}

\maketitle
\begin{abstract}
This work applies resolvent analysis to incompressible flow through a rectangular duct, in order to identify dominant linear energy-amplification mechanisms present in such flows. In particular, we formulate the resolvent operator from linearizing the Navier--Stokes equations about a two-dimensional base/mean flow. The laminar base flow only has a nonzero streamwise velocity component, while the turbulent case exhibits a secondary mean flow (Prandtl’s secondary flow of the second kind). A singular value decomposition of the resolvent operator allows for the identification of structures corresponding to maximal energy amplification, for specified streamwise wavenumbers and temporal frequencies. Resolvent analysis has been fruitful for analysis of wall-bounded flows with spanwise homogeneity, and here we aim to explore how such methods and findings can extend for a flow in spatial domains of finite spanwise extent. We investigate how linear energy-amplification mechanisms (in particular the response to harmonic forcing) change in magnitude and structure as the aspect ratio (defined as the duct width divided by its height) varies between one (a square duct) and ten. 
We additionally study the effect that  secondary flow has on linear energy-amplification mechanisms, finding that in different regimes it can either enhance or suppress amplification. We further investigate how the secondary flow alters the forcing and response mode shapes leading to maximal linear energy amplification.


\end{abstract}

\section*{Nomenclature}
{\renewcommand\arraystretch{1.0}
\noindent\begin{longtable*}{@{}l @{\quad=\quad} l@{}}
$x$ & streamwise coordinate\\
$y,z$ & wall-normal and spanwise coordinates\\
$A$ & duct aspect ratio, and normalized channel width \\
$k_x$ & wavenumber in the streamwise direction \\
$k_z$ & wavenumber in the spanwise direction \\
$\mathcal{L}$ & linearized equations \\
$\omega$ & complex temporal frequency \\
$\mathcal{H}_\omega$ & resolvent operator \\
$\sigma$ & singular value, amplification level or resolvent gain\\
$\bsq$ & state vector \\
$\bsf$ & forcing vector \\
$\bsu$ & velocity field $(u,v,w)$ \\
$\bsu'$ & fluctuating field \\
$\bsU$ & mean velocity field $(U,V,W)$ \\
$P, p'$ & mean and fluctuating pressure \\
$\psi$ & resolvent response mode  \\
$\phi$ & resolvent forcing mode\\
$\Lambda_\epsilon$ & $\epsilon$-pseudospectrum \\
$Re$ & Reynolds number \\
\end{longtable*}}

\section{Introduction}
\lettrine{P}{lanar} wall-bounded flows are among the most well-understood, owing to their geometrical and mathematical simplicity, and to the wide range of studies dedicated to such configurations. As a highly non-comprehensive set of examples, such investigations range from classical theoretical analysis of underlying operators \cite{drazin2004hydrodynamic,lin1955hydrodynamic} and more recent analysis of linear amplification \cite{jovanovic2005componentwise,reddy1993energy,butler1992optimal,sharma2013resolvent,moarref2013channels}, to a wide range of numerical investigations of turbulent channel flows \cite{moin1982numerical,moser1999direct, del2004scaling,lee2015direct}.

While channel flow serves as a productive testbed for studying the properties of canonical shear flows, and while certain characteristics of wall-bounded turbulent shear flows appear to be ubiquitous across a range of systems, it is important to understand how, where, and why realistic flows depart from such ideal configurations.  An obvious point of departure between channel (Poiseuille) flow and real-world examples is the fact that truly spanwise-homogeneous flows do not exist in reality. In practice, flow through ducts with rectangular cross sections are ubiquitous across a range of applications, including cooling and ventilation systems.

It has been known for the best part of a century that turbulent flow through rectangular ducts generates secondary mean flow \cite{nikuradse1930,prandtl1931} (Prandtl's secondary flows of the second kind), comprising of counter-rotating vortices that transport streamwise momentum towards the duct corners.  See Ref.~\cite{bradshaw1987turbulent} for a more general review of turbulent secondary flows. While the existence of such secondary flows has been long-known, a full understanding of finite-aspect-ratio effects on duct flows remains an active area of study.  Indeed, simply obtaining accurate characterization of mean secondary flows can require averaging over a large period of time, on the order of 3,000 convective time units  \cite{vinuesa2018}.
More broadly, a range of contemporary studies have considered, for example, numerical \cite{orlandi2019} and experimental \cite{owolabi2016experiments} studies of transition in ducts, and studies of  incompressible \cite{vinuesa2014,vinuesa2015minimum,vinuesa2018,orlandi2019} and compressible \cite{huang1995compressible} turbulent duct flows. 
Ref.~\cite{orlandi2019} found critical Reynolds number for transition was insensitive to aspect ratios greater than unity, 
while Ref.~\cite{vinuesa2015minimum} argues that duct flows can require aspect ratios of at least 24 to be comparable to spanwise-periodic channel flow configurations.
Ref.~\cite{matin2018coherent} utilizes a proper orthogonal to identify energetic coherent structures in turbulent duct flows, while the time-evolving dynamics of such structures are similarly studied in \cite{khan2020dynamics}.
In Ref.~\cite{zaripov2021mechanism}, mechanisms leading to transient near-wall reverse flow events in a square duct are proposed.
 
The resolvent formulation of the Navier--Stokes equations \cite{mckeon2010resolvent,mckeon2017engine} has proven to be a highly-useful framework for understanding and modeling the dynamics of wall-bounded and shear-driven flows, such as in the prediction and modeling of coherent structures \cite{sharma2013resolvent,abreu2020spectral} and statistics \cite{moarref2013channels,mcmullen2020interaction,towne2020resolvent}. While resolvent analysis has been applied to a broad range of flows (including cavity  \cite{gomez2016reduced,Qadri:PRF17},  airfoil \cite{thomareis2018resolvent,symon2019tale,yeh2019resolvent,yeh2020resolvent}, and jet \cite{garnaud13jet,jeun2016jet,towne2018spectral,pickering2021resolvent,sun2020resolvent} flows), there has been comparatively few investigations considering resolvent analysis of flows that are not spanwise-homogeneous or axisymmetric. 
 
While we are not aware of prior works performing resolvent analysis of rectangular duct flows, asymptotic linear stability analysis has been previously considered for such flows in Refs.~\cite{tatsumi1990,theofilis2004stability}. Optimal transient energy grow analysis for square duct flow has also been investigated in Ref.~\cite{biau2008transition}, where it is found that  optimal transient energy growth trajectories for square duct flows often do not trigger transition.
 
This paper proceeds as follows. Sect.~\ref{sec:method} first describes the resolvent analysis formulation and mean/base flows used for the rectangular duct configuration. Following this, in Sect.~\ref{sec:results} we validate the numerical methods against past studies, before presenting results of applying resolvent analysis to (square and rectangular) laminar and turbulent duct flows. Sect.~\ref{sec:disc} summarizes findings, and discusses potential future steps.



\section{Methodology}
\label{sec:method}
 In this section, we  describe the  resolvent formulation of the governing equations, which assumes a Fourier decomposition in both the streamwise spatial direction ($x$), and in time. We also describe the general aspects of the duct-flow geometry, and present  analytic expressions for the laminar equilibrium velocity profile for pressure-driven duct flow. 

\subsection{Pseudospectral analysis of a linear operator}
We begin by considering a dynamical system governed by 
\begin{equation}
\label{eq:dynsys}
    \dot{\bsq}(\bsx,t)+ \mathcal{L} \bsq(\bsx,t) = \bsf(\bsx,t),
\end{equation}
where $\bsq$ denotes the state of the system with respect to a reference state $\bsq_0$, $\mathcal{L}$ is a linear operator and $\bsf$ represents an input to the system. Assuming that an instantaneous state can be decomposed into spatial and temporal (harmonic) terms allows us to assume solutions of the form
\begin{equation}
    \bsq(\bsx,t)= \hat{\bsq}(\bsx)\exp{(-{i}\omega t)},
\end{equation}
where $\omega$ is a complex number. In the cases where the forcing is nonexistent, equation \ref{eq:dynsys} becomes the \textit{eigenproblem} $(-{i}\omega)\hat{\bsq}+\mathcal{L}\hat{\bsq}=0$, in which the term ${i}\omega$ is an eigenvalue of $\mathcal{L}$.
However, if a forcing term $\bsf$ exists, both the properties of the resolvent operator and the nature of $\bsf$ will influence the response of the system. In particular, assuming that this forcing term can be written as a Fourier mode in time
\begin{equation}
    \bsf(\bsx,t)= \hat{\bsf}(\bsx)\exp{(-{i}\omega t)},
\end{equation}
allows us to eliminate the temporal dependence of the terms in equation \ref{eq:dynsys}, and to recast the governing equation for the state of the spatial component $\hat{\bsq}$ as
\begin{equation}
\label{eq:govResolvent}
    (-{i}\omega\textbf{I}+\mathcal{L}) \hat{\bsq}(\bsx) = \hat{\bsf}(\bsx).
\end{equation}
This provides a formulation that maps an exogenous disturbance, represented by the forcing term $\hat\bsf$, to the state of the system $\hat{\bsq}$ , by
\begin{equation}
\label{eq:res}
    \hat{\bsq} =(-{i}\omega\textbf{I}+\mathcal{L})^{-1} \hat{\bsf} := \mathcal{H}_\omega \hat{\bsf},
\end{equation}
where  $\mathcal{H}_\omega$ is the resolvent operator at a specified frequency $\omega$ (provided that ${i}\omega$ is not in the spectrum of $\mathcal{L}$).

Here, we are interested in studying forcing and response pairs $\hat{\bsf}$ and $\hat{\bsq}$ that are related via Eq.~\ref{eq:res}, particularly in cases where a small forcing $\hat{\bsf}$ can induce a large response in $\hat{\bsq}$. This is revealed by pseudospectral analysis of the resolvent operator $\mathcal{H}_\omega$. The singular value decomposition (SVD) of the resolvent operator yields
\begin{equation}
    \mathcal{H}_\omega=\sum_{j=1}^\infty \psi_j \sigma_j \phi^*_j,
\end{equation}
where $\sigma_k>\sigma_{k+1}\geq 0$ for all $k$. 
Note in particular that the leading singular value and singular vectors satisfy
\begin{equation}
    \sigma_1=\max_{||\phi=1||} ||\mathcal{H}_\omega \phi||,
\end{equation}
and
\begin{equation}
    \phi_1=\argmax_{||\phi=1||} ||\mathcal{H}_\omega \phi||,
\end{equation}
with 
\begin{equation}
    \psi_1=\sigma_1^{-1} \mathcal{H}_\omega \phi_1.
\end{equation}
The leading singular value $\sigma_1$ of the resolvent operator $\mathcal{H}_\omega$ is directly related to the $\epsilon$-pseudospectra ($\epsilon>0$) of a linear operator $\mathcal{L}$, defined as the region of the complex plane $\Lambda_\epsilon \subset \mathbb{C}$ that satisfies
\begin{equation}
    \Lambda_\epsilon(\mathcal{L})=\lbrace z: (\mathcal{L+E})\textbf{\textit{u}} = z \textbf{\textit{u}}, \textnormal{for some $\textbf{\textit{u}}$ and $\mathcal{E}$, with $\|\mathcal{E}\| \leq \epsilon$} \rbrace,
\end{equation}
where the operator  $\mathcal{E}$  maps between the same spaces as $\mathcal{L}$.
In particular, for any $z \in \mathbb{C}$, the leading singular value $\sigma_1$ of the resolvent can be defined as $\sigma_1^{-1}=\min \lbrace \epsilon: z \in \Lambda_\epsilon (\mathcal{L})\rbrace$.


If a linear operator $\mathcal{L}$ is normal (i.e.~if  $\mathcal{L}\mathcal{L}^*=\mathcal{L}^*\mathcal{L}$), then the $\epsilon$-pseudospectra is constituted by the locations in the complex plane $\mathbb{C}$ that are within a radius $\epsilon$ from the spectrum of $\mathcal{L}$. In this work, we will be concerned with nonnormal operators, for which the $\epsilon$-pseudospectra can be much larger, meaning that even frequencies that are far from eigenvalues can be highly amplified by the linearized dynamics.

\subsection{The incompressible Navier--Stokes equations in resolvent formulation}
Consider the non-dimensionalized incompressible Navier--Stokes equations 
\begin{equation}
\label{eq:momentum}
\frac{\partial \bsu}{\partial t}+(\bsu \cdot \nabla) \bsu = -\nabla p+\frac{1}{Re}\Delta \bsu,
\end{equation}
\begin{equation}
\label{eq:continuity}
\nabla \cdot \bsu =0,
\end{equation}
where $\bsu(\bsx,t)=
\left(u(\bsx,t),v(\bsx,t),w(\bsx,t)\right)$ represent the velocity field and $p(\bsx,t)$ denotes pressure. Writing the instantaneous state as the sum of the mean and the fluctuating parts, respectively, gives $\bsu=\bsU+\bsu'$ and $p=P+p'$, assuming that the dynamics are statistically stationary. Introducing this decomposition into Eq.~\eqref{eq:momentum} and expanding the terms we obtain 
\begin{equation}
\label{eq:momentum2}
\frac{\partial \bsu'}{\partial t}+ \bsU \cdot \nabla \bsU+ \bsU\cdot \nabla \bsu'+\bsu'\cdot \nabla\bsU+\bsu' \cdot \nabla \bsu'=-\nabla P -\nabla p' +\frac{1}{Re}\left(\Delta\bsU +\Delta \bsu' \right).
\end{equation}
Now, if we evaluate the time-average of the equation above, we obtain
\begin{equation}
\label{eq:mean}
\bsU \cdot \nabla \bsU+ \overline{\bsu' \cdot \nabla \bsu'}=-\nabla P  +\frac{1}{Re}\Delta\bsU.
\end{equation}
Subtracting Eq.~\eqref{eq:mean} from Eq.~\eqref{eq:momentum2} gives
\begin{equation}
\label{eq:fluct}
\frac{\partial \bsu'}{\partial t}+ \bsU\cdot \nabla \bsu'+\bsu'\cdot \nabla\bsU+\nabla p'-\frac{1}{Re}\Delta \bsu'=-\bsu' \cdot \nabla \bsu' -\overline{\bsu' \cdot \nabla \bsu'}.
\end{equation}
These equations are closed with the  continuity equation for the fluctuating components,
\begin{equation}
\label{eq:continuity_fluct}
\nabla \cdot \bsu' =0.
\end{equation}
We consider systems which are statistically stationary in time and homogeneous in the streamwise direction, allowing us to consider trajectories for both $\bsu'$ and $p'$ of the form
\begin{equation}
\label{eq:udecomp}
\bsu' (\bsx,t) = \hat{\bsu}(y,z) \exp{[{i}(k_x x-\omega t)]},
\end{equation}
\begin{equation}
\label{eq:pdecomp}
p' (\bsx,t) = \hat{p}(y,z) \exp{[{i}(k_x x-\omega t)]},
\end{equation}
i.e.~assuming Fourier modes for the spatial structures traveling in the streamwise direction and the temporal dimension. We can substitute Eqs.~\eqref{eq:udecomp}-\eqref{eq:pdecomp} into Eqs.~\eqref{eq:fluct}-\eqref{eq:continuity_fluct}, which allows us to recast the governing equations in terms of the formulation given in Eq.~\eqref{eq:res}, giving
 \begin{equation}
 {
\label{eq:resolvent}
\left[-i\omega\textbf{I}-
\begin{pmatrix}
L & \partial_yU & \partial_zU & {i}k_x{I} \\
0 & L+\partial_yV & \partial_zV & \partial_y \\
0 & \partial_yW & L+\partial_zW & \partial_z \\
{i}k_x{I} & \partial_y & \partial_z & 0 
\end{pmatrix}\right]
\begin{pmatrix}
\hat{u} \\
\hat{v} \\
\hat{w} \\
\hat{p}
\end{pmatrix}=
\mathcal{H}_\omega^{-1}
\begin{pmatrix}
\hat{u} \\
\hat{v} \\
\hat{w} \\
\hat{p}
\end{pmatrix}
=
\begin{pmatrix}
\hat{f}_u \\
\hat{f}_v \\
\hat{f}_w \\
\hat{f}_p 
\end{pmatrix}, }
\end{equation} 
where $L = {i}k_xU+V \partial_y+W\partial_z-\frac{1}{Re}\Delta$,
and $(U,V,W) = \bsU$ are the components of the mean velocity.
Here, the nonlinear terms on the right-hand side have been replaced by the forcing term, $\hat\bsf$.

\subsection{Description of two-dimensional duct flows}
We consider flows that are homogeneous in the streamwise direction ($x$), while the other two dimensions ($y$,$z$) are confined to the domain $\Omega=\lbrace y \in [-1,1]\rbrace \times \lbrace z \in [-A,A]\rbrace$, with $A$ being the aspect ratio of the rectangular cross-section of the duct. 
 No-slip and no-penetration conditions are imposed at all wall-boundaries. 
 

For laminar flow, there is an equilibrium velocity profile with only a velocity component in the $x$ direction.  
It is possible to determine this velocity profile $\bsU(y,z) = (U(y,z),0,0)$ analytically, in the form of infinite series. 
 First, a steady laminar flow  through a rectangular duct driven by a constant pressure gradient along the streamwise direction satisfies Poisson's equation
\begin{equation}
\label{eq:poisson}
\Delta U(y,z)= C 
\end{equation}
with boundary conditions
\begin{equation}
\bsU(y=-1,z)=\bsU(y=1,z)=\bsU(y,z=-A)=\bsU(y,z=A)= {\bf 0}.
\end{equation}
Taking $C=-2$ and scaling the domain to comply with the boundaries described above, Eq.~\eqref{eq:poisson} has a solution of the form (e.g. Ref.~\cite{panton2013incompressible})
\begin{equation}
\label{eq:pressuredriven}
U(y,z)=1-y^2-4\left(\frac{2}{\pi}\right)^3 \sum_{n=0}^\infty \frac{(-1)^n}{(2n+1)^3} \frac{\cosh[(2n+1)\pi z/2]\cos[(2n+1)\pi y/2]}{\cosh[(2n+1)\pi A/2]}.
\end{equation}
Note that in the limit $A \to \infty$,  this solution converges to standard planar Poiseuille flow, with  $U(y)=1-y^2$.


\begin{figure}[ht!]
\vspace*{-0.3cm}
\centering {
{\hspace*{-1.6cm}\includegraphics[width= 1.2\textwidth]{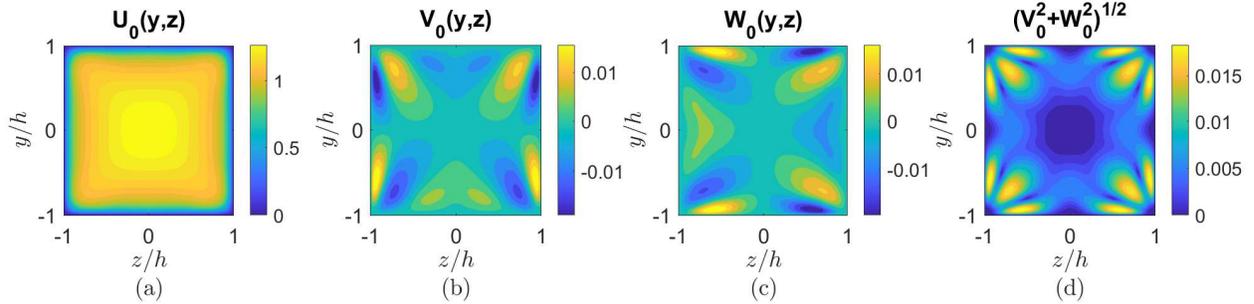}} }
\vspace*{-0.5cm}
\caption{Components of the turbulent mean velocity profile for flow through a square duct with Reynolds number $Re_b=2500$ \cite{vinuesa2018}.
}
\label{fig:turbulentMean}
\end{figure}

We also wish to examine the contribution of secondary flows to the linear amplification mechanisms for this configuration. In this case, instead of using an exact solution, we utilize mean velocity profiles obtained from the direct numerical simulations  described in \citet{vinuesa2018}, an example of which is shown in  Fig.~\ref{fig:turbulentMean}). Notice how despite the small magnitude of the secondary motions, the isovelocity contours shown in Fig.~\ref{fig:turbulentMean}(a) are non-convex \cite{prandtl1931}. We will consider  data for aspect ratios $A$ between 1 and 10, and a Reynolds number (based on the bulk velocity and channel half-height) of $Re_b=2500$. This corresponds to a friction Reynolds number (at the centerline) of $Re_\tau \approx 180$. 

\section{Results}
\label{sec:results}
This section first presents a validation of the numerical methods in Sect.~\ref{sec:numerical}, through a comparison with prior stability analyses results for laminar duct flows with varying aspect ratios.
Sects.~\ref{sec:analysisRatios} and \ref{sec:analysisCritL} present the central findings of this work. Both sections consider resolvent analysis for three base/mean profile types: laminar flow, turbulent flow, and turbulent flow where the secondary flow components are removed. Sect.~\ref{sec:analysisRatios} focuses on the effect of the duct aspect ratio, while Sect.~\ref{sec:analysisCritL} examines the square duct case in more detail, examining the differences between these three profile types as a function of streamwise wavenumber and temporal frequency.


\subsection{Numerical Formulation and Validation}
\label{sec:numerical}
The operator described in Eq.~\eqref{eq:resolvent} is formulated and discretized using a Chebyshev collocation method in both the $y$ and $z$ directions. Before performing resolvent analysis, the code is validated by comparing the spectra (i.e., asymptotic stability analysis, solutions to $\mathcal{L}\bsq = i\omega \bsq$) of the linearized Navier-Stokes operator for a pressure-driven rectangular duct flow with the results of previous studies of \citet{theofilis2004stability} and \citet{tatsumi1990}. 

\begin{figure}[ht!]
\vspace{-0.4cm}
\centering{
{\includegraphics[width= 1\textwidth]{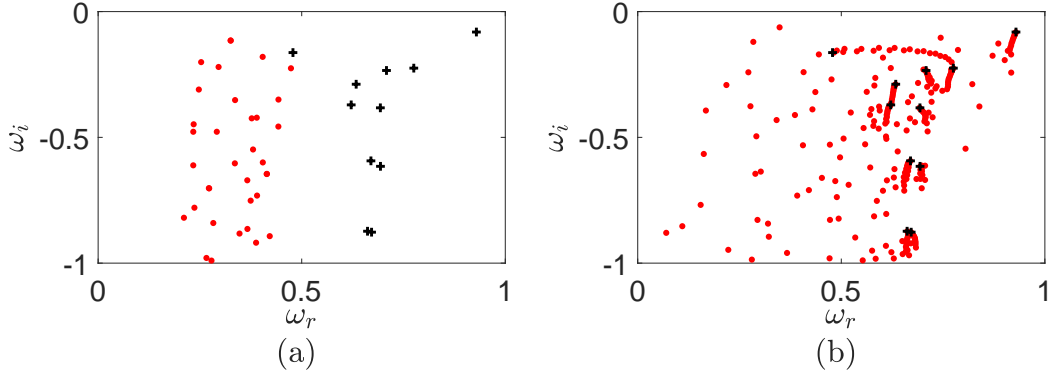}} }
\vspace{-0.5cm}
\caption{Least-stable eigenvalues of the linearized Navier--Stokes equations for a laminar pressure-driven square duct flow (red) with $A=1$ (a) and $A=10$ (b) using a spatial grid $(N_y,N_z)=(64,32)$, along with the eigenvalues of the resolvent in planar Poiseuille flow (black) in a channel with a spanwise wavenumber $k_z=0$, with $Re=100$, $k_x=1$.
}
\label{fig:stabilityPPF}
\end{figure}

Sample validation results at various Reynolds numbers and aspect ratio are shown in Table \ref{tab:mode1Theofilis}. In these cases, the spatial grid is formed using $(N_y,N_z)=(80,100)$ collocation points in the $y$ and $z$ directions.  It is observed that the identified critical frequencies $\omega_r = \text{Real}(\omega)$ are generally in agreement with the literature, agreeing to three significant figures for $Re\leq 10,400$. 
\begin{table}[t]
\small
\begin{tabular}{|c c c c c|} 
 \hline
 A & $Re$ & $k_x$ & $\omega_r$ & $\omega_r$, current  \\
 \hline\hline
 3.5 & 36,600 & 0.71 & 0.12353 & 0.128322 \\ 
 \hline
 4.0 & 18,400 & 0.80 & 0.16187 & 0.164351 \\
 \hline
 5.0 & 10,400 & 0.91 & 0.21167 & 0.211952 \\
 \hline
 6.0 & 8,200 & 0.94 & 0.22925 & 0.229307\\
 \hline
 8.0 & 6,800 & 0.98 & 0.24963 & 0.249874\\  
 \hline
 25.0 & 5,772 & 1.02 & 0.26960 & 0.26975\\  
 \hline
\end{tabular}
\caption{\label{tab:mode1Theofilis}Comparison between critical frequencies $\omega_r$ of the least stable eigenvalue of the linearized Navier--Stokes equations, identified using our code (current) and Ref.~\cite{theofilis2004stability}). Comparisons are made for various aspect ratios, $A$, Reynolds numbers, $Re$, and streamwise wavenumbers, $k_x$.}
\end{table}
 We also show eigenvalues in a region near the marginal stability axis in Fig.~\ref{fig:stabilityPPF} for cases with $A = 1$ and 10, in comparison with those of plane Poiseuille flow (i.e. $A \to \infty$), with $k_z=0$. It can be seen that as the aspect ratio increases, the spectrum includes clusters of eigenvalues that approach those for an infinitely-wide spatial domain. 

\subsection{Aspect ratio effects on resolvent analysis  of laminar and turbulent rectangular duct flows}
\label{sec:analysisRatios}
This section studies the influence of the aspect ratio on leading resolvent resolvent modes and amplification levels for laminar and turbulent  duct flows. For this section, we consider only one streamwise  wavenumber, $k_x$, and wave speed, $c = \omega/k_x$. 
Note that all volume flow rates have been normalized to one, enabling direct comparisons between results.

\begin{figure}[ht!]
\centering {
{\hspace*{-2.1cm}\includegraphics[width= 1.25\textwidth]{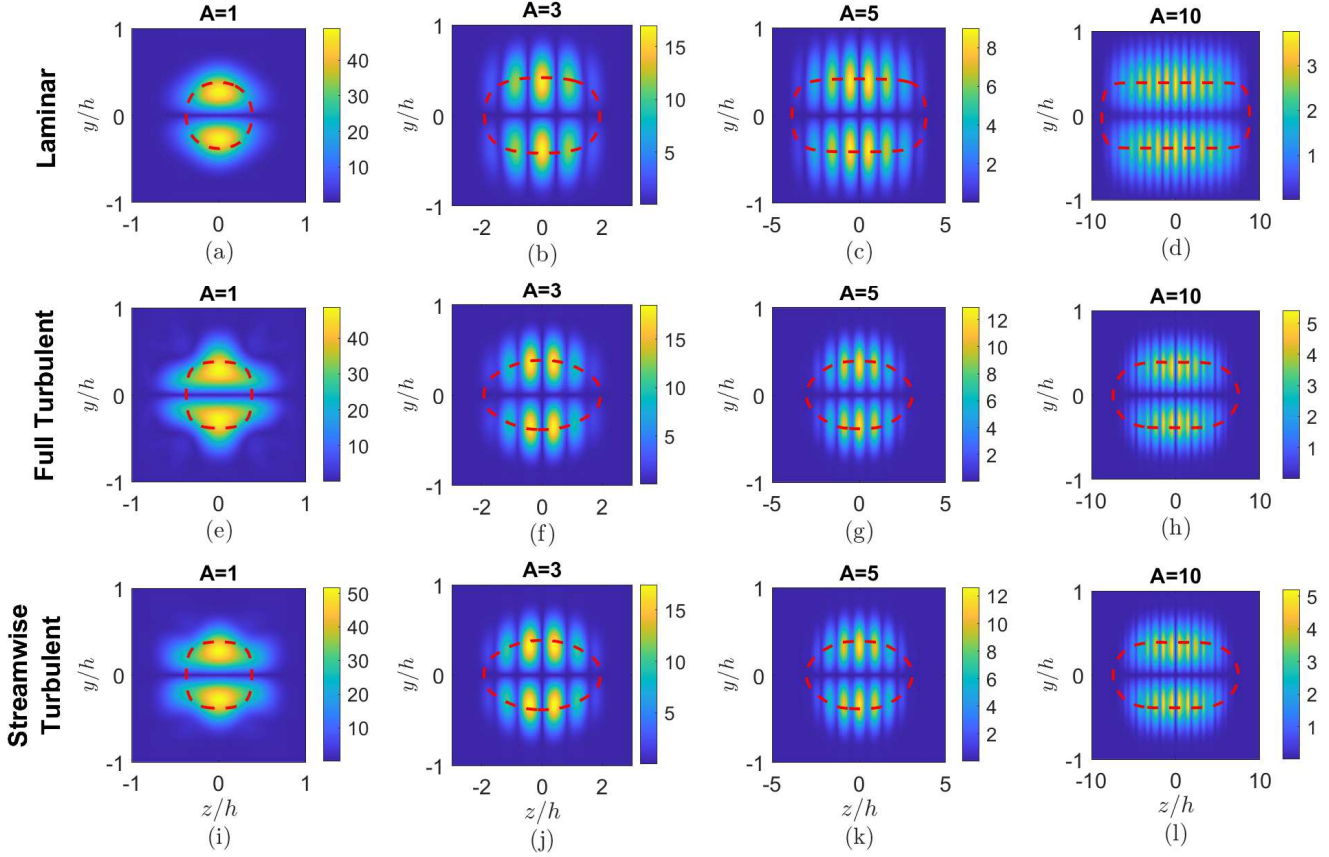}} }
\vspace{-0.8cm}
\caption{Amplitude of the leading resolvent response modes in the streamwise direction $\psi_u(y,z)$ for a laminar (a)-(d), turbulent (e)-(h) and streamwise turbulent (i)-(l), rectangular duct flow with varying aspect ratios $A\in\lbrace 1,2,5,10\rbrace$, with $Re_b=2500$, $k_x=1$ and $c=1.25$. Critical-layer locations indicated by red dashed lines.  Note that plots with $A >1$ are not drawn to scale.
}
\label{fig:ResponseModes_ratios}
\end{figure}

In Fig.~\ref{fig:ResponseModes_ratios} we present the leading response modes in the streamwise direction $\psi_u(y,z)$ for  aspect ratios $A=\in\{1,2,5,10\}$ with $Re_b=2500$, $k_x=1$ and $c=1.25$ for the laminar case, the turbulent case, and the turbulent case with only the streamwise velocity component. It is observed that all modes are concentrated about the critical layers (the locations where the wavespeed matches the value of the streamwise component of the mean, indicated with red dashed lines on the plots). The presence of the wall-boundaries forces the response modes to be localized in the inner region of the domain for the smallest aspect ratio, $A=1$. This is primarily a result of the critical layer being closer to the center of the domain in this case. In general, it seems that the differences in the mode shapes between the laminar and turbulent cases can largely be attributed to the differences in critical layer location. In Sect.~\ref{sec:analysisCritL}, we will consider adjustments the wavespeed to enable a more direct comparison between the laminar and turbulent cases.  
In the laminar case in particular, we observe modes that become concentrated at a constant distance from the upper and lower boundaries for higher ratios, with periodic oscillations in the spanwise ($z$) direction. 
For the larger aspect ratio cases, it is tempting to make a direct comparison with the case of an infinite aspect ratio with Fourier decomposition in the spanwise direction, though note that Fourier modes would not have the oscillations in their amplitude that are observed here (they would have a constant amplitude in the spanwise direction, with oscillations only in the real and imaginary components). Note lastly that the different scales of mode amplitude are primarily due to the domains being different sizes (as all modes normalized to have unit energy). 

\begin{figure}[ht!]
\vspace{0cm}
\centering {
{\hspace*{-1.3cm}\includegraphics[width=1.15\textwidth]{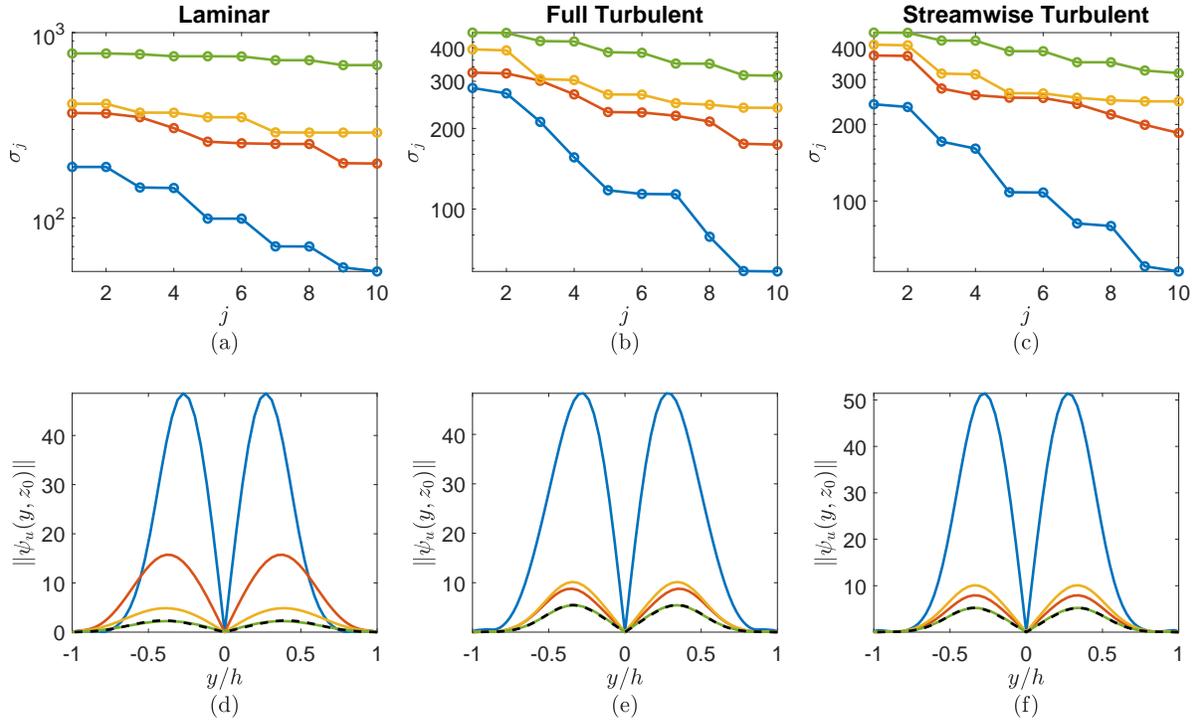}} }
\vspace{-0.4cm}
\caption{(a)-(c) First normalized resolvent gains and (d)-(f) planar projection of maximal amplitude of the response modes in figure \ref{fig:ResponseModes_ratios} for laminar, turbulent and streamwise turbulent, pressure-driven rectangular duct flows with varying aspect ratios $A=1$ (blue), $A=3$ (orange), $A=5$ (yellow), $A=10$ (green), and planar Poiseuille flow (\protect\blackDashed) for laminar and turbulent channel flow (\protect\blackDashed), in a channel with a spanwise number $k_z=2\pi$, with $Re_b=2500$, $k_x=1$ and $c=1.25$.
}
\label{fig:SingVals_ratios}
\end{figure}

Fig.~\ref{fig:SingVals_ratios}(a)-(c) compares the amplification levels (singular values) of the leading 10 modes for the cases depicted in Fig.~\ref{fig:ResponseModes_ratios}. 
We find that the leading singular values decrease with aspect ratio for all cases, with the most notable drop off occurring for the laminar case between $A=2$ and $1$. We further observe that the singular values decay more rapidly for the $A=1$ case for all mean profiles. In general, however, the singular values decay relatively slowly, which 
is in contrast to what is typically observed for the spanwise-homogeneous case, which 
usually features a sharp drop off after the first two singular values (at a specified spanwise wavelength).  This slow decay is perhaps expected, since for large aspect ratios the two-dimensional analysis includes all possible spanwise fluctuations, which are decoupled into separate operators for the spanwise-homogeneous case. Nevertheless, these findings have implications for the use of such analysis for accurate representation and prediction of unsteady structures in such flows using a small number of modes. 

Note also that while the response modes in Fig.~\ref{fig:ResponseModes_ratios} appear to be the same for the full and streamwise-only turbulent cases, Figs.~\ref{fig:SingVals_ratios}(b)-(c) show notable differences in the corresponding singular values.  In particular, the presence of secondary mean flow leads to an increase in the leading singular value for $A=1$, but a decrease for $A=3$ in comparison to the streamwise-only case.  This suggests that the secondary flows have a non-negligible influence on linear amplification, which will be studied in more detail in Sect.~\ref{sec:analysisCritL}.

Cross sections at a $z$-location corresponding to maximal mode amplitude are shown in Fig.~\ref{fig:SingVals_ratios}(d)-(f). We again observe similar behavior for aspect ratios > 1 in both laminar and turbulent cases. We also show that the leading mode amplitudes are in very close agreement to that for a homogeneous spanwise direction (here plotted with a spanwise wavenumber $k_z = 2\pi$). This similarity suggests that the dominant linear amplification mechanisms for these cases are indeed similar to those obtained for the laminar and turbulent spanwise-homogeneous system. 

%
\subsection{The influence of secondary flows on linear amplification in square ducts}
\label{sec:analysisCritL}
\begin{figure}[ht!]
\centering {
{\hspace*{0cm}\includegraphics[width= 0.5\textwidth]{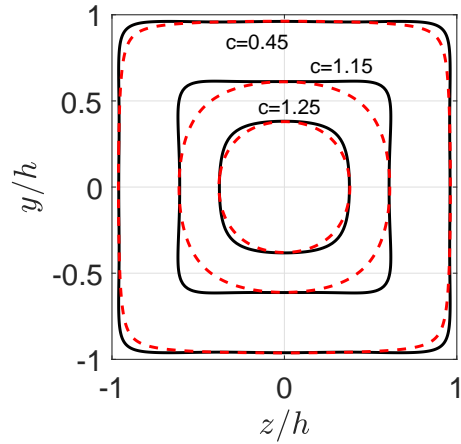}} }
\vspace{0cm}
\caption{Critical-layer locations that will be used in Figs.~\ref{fig:singVals_critLayer}-~\ref{fig:fullModes_All_kx10c1p25} for the full turbulent system (black) and the equivalent in the laminar system (dashed red). Note that the stated wavespeeds apply to the turbulent mean cases.
}
\label{fig:critLayers}
\end{figure}

We now extend the previous analysis across a range of wavespeeds, $c$, and the streamwise wavenumbers, $k_x$, focusing on the square duct ($A=1$) case, where secondary flows should have the largest influence. 
We are particularly interested in determining the extent to which secondary flows directly influence linear amplification mechanisms. 
Additionally, direct comparison with the laminar case might help understand the effect of the mean streamwise velocity distribution. To enable a more direct comparison between the laminar and turbulent cases, in this section we adjust the wavespeed used for the laminar mean cases so that the critical layer locations are similar to the turbulent cases. 
Fig.~\ref{fig:critLayers} shows this alignment of the laminar and turbulent critical layers.
 Henceforth, we will refer to the wavespeeds used for the turbulent mean cases, with the understanding that the wavespeeds for the laminar cases have been adjusted to best match these critical layer locations.

\begin{figure}[ht!]
\centering {
{\hspace*{0cm}\includegraphics[width= 1\textwidth]{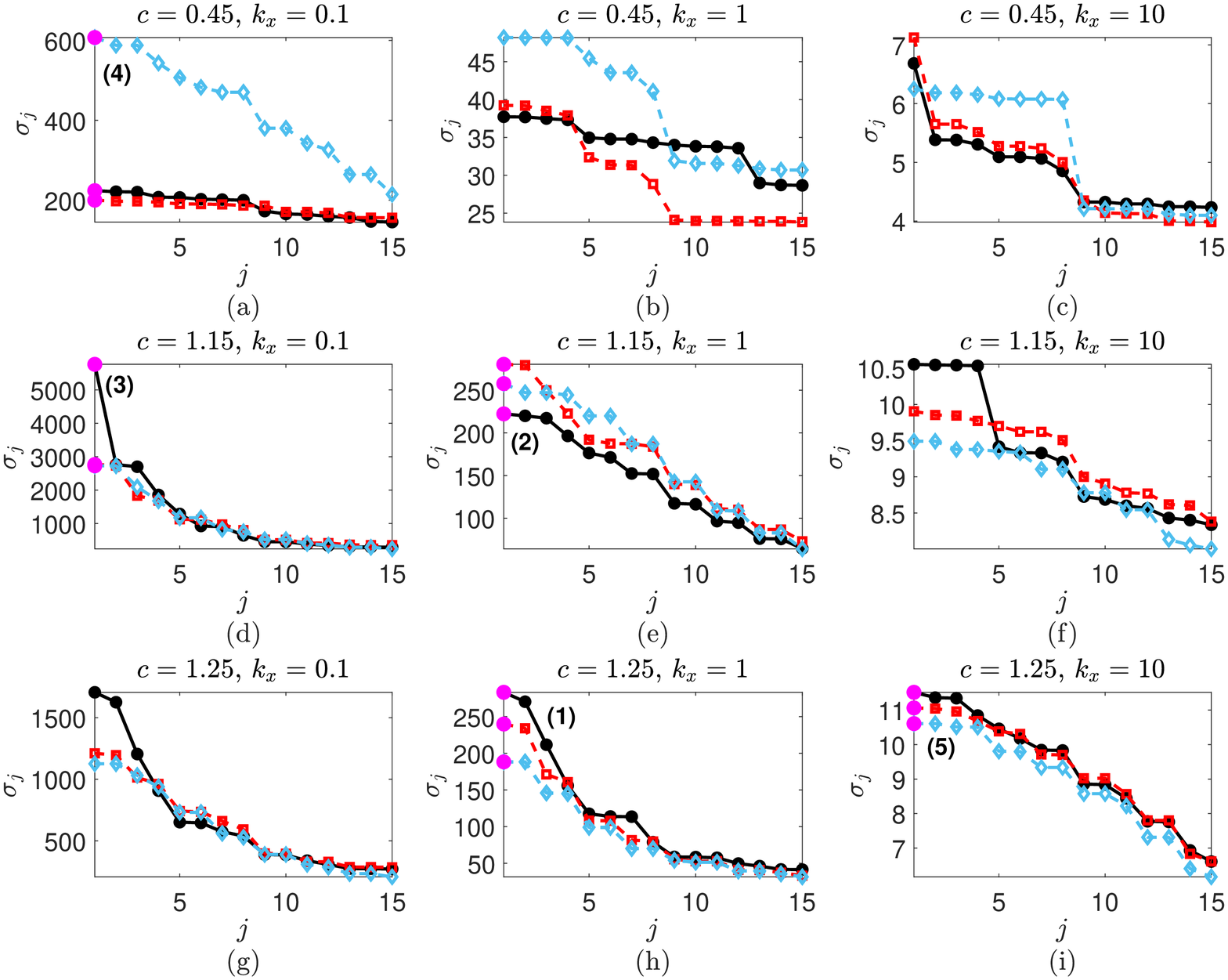}} }
\vspace{-1cm}
\caption{Singular values of the full turbulent (black), the streamwise component (red) of the turbulent, and laminar (blue) square duct flow with $Re_b=2500$ over a range of critical layer locations $c=\{0.45,1.15,1.25\}$ and streamwise wavenumbers $k_x=\{0.1,1,10\}$. Labels (1)-(5) refer to the resolvent modes depicted in Figs.~\ref{fig:fullModes_All_kx1c1p25}-~\ref{fig:fullModes_All_kx10c1p25}, respectively, with pink markers specifying modes plotted in these figures.
}
\label{fig:singVals_critLayer}
\end{figure}

In Fig.~\ref{fig:singVals_critLayer} we show the singular values of the laminar and turbulent systems (with and without secondary flows) over a range of wavespeeds $c \in \{0.45,1.15,1.25\}$ and streamwise wavenumbers $k_x \in \{0.1,1,10\}$. Perhaps the most notable observation is that there are parameters for which the presence of secondary flows enhances amplification, and those which reduce the amount of amplification. Aside from low-speed modes ($c = 0.45$) with $k = 0.1$ and 1, the laminar cases also appear to follow similar behavior to the turbulent cases. The remainder of this section will focus on studying the forcing and response modes for several of the cases shown in Fig.~\ref{fig:singVals_critLayer}, as indicated by pink markers.

\begin{figure}[ht!]
\centering {
\vspace{-0.5cm}
{\hspace*{-2.4cm}\includegraphics[width= 1.3\textwidth]{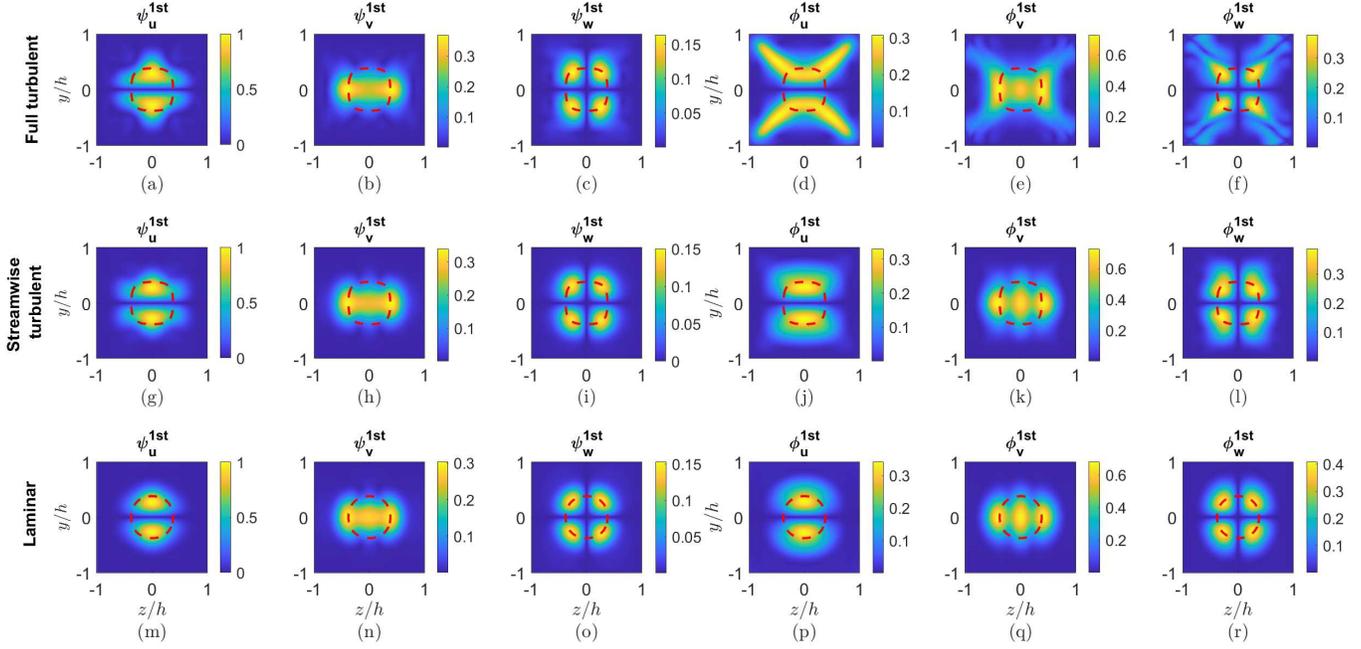}} }
\vspace{-0.6cm}
\caption{Normalized amplitude of the optimal response $\psi(y,z)$ and forcing $\phi(y,z)$ modes of the resolvent operator for a laminar pressure-driven (a)-(f), full turbulent (g)-(l) and streamwise component of a turbulent (m)-(r), square duct flow with with $Re_b=2500$, $k_x=1$ and $c=1.25$. Critical-layer locations indicated by red dashed lines.
}
\label{fig:fullModes_All_kx1c1p25}
\end{figure}

In Fig.~\ref{fig:fullModes_All_kx1c1p25} we compare the resolvent modes of the full and streamwise turbulent flows, as well as the laminar case, for $k_x=1$ and $c=1.25$. We first note that all components of the response modes $\psi(y,z)$  are very similar. In all cases, the streamwise component dominates the response, while the $v$ component of the forcing is largest. Each forcing mode component for the laminar and streamwise-only turbulent cases are very similar, however there is a notable difference for the forcing modes for the turbulent case with secondary flows present. The forcing modes (particularly the streamwise velocity component) extend much further towards the corners, where the secondary flows are largest in magnitude (see Fig.~\ref{fig:turbulentMean}(d)). This highlights the effect that small secondary flows can have in altering linear amplification mechanisms (in this case enhancing the amplification, per Fig,~\ref{fig:singVals_critLayer}(h)). Furthermore, these modes suggest that secondary flows can facilitate larger spatial separation between locations of forcing and response.

\begin{figure}[ht!]
\centering {
\vspace{-0.45cm}
{\hspace*{-2.4cm}\includegraphics[width= 1.3\textwidth]{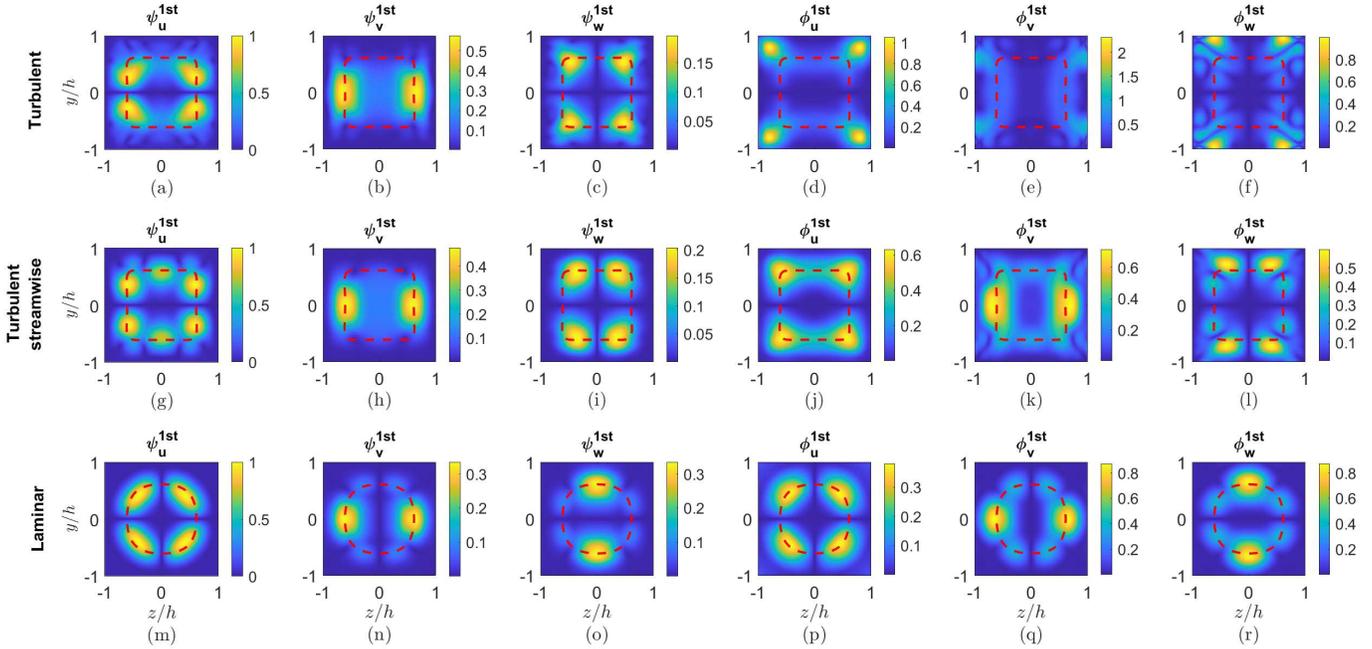}} }
\vspace{-0.6cm}
\caption{Normalized amplitude of the optimal response $\psi(y,z)$ and forcing $\phi(y,z)$ modes of the resolvent operator for a laminar pressure-driven (a)-(f), full turbulent (g)-(l) and streamwise component of a turbulent (m)-(r), square duct flow with with $Re_b=2500$, $k_x=1$ and $c=1.15$. Critical-layer locations indicated by red dashed lines.
}
\label{fig:fullModes_All_kx1c1p15}
\end{figure}

Fig.~\ref{fig:fullModes_All_kx1c1p15} represents a case in which the resolvent modes bear similarity across all three systems, though with some notable differences. In particular, according to the amplification levels shown in Fig.~\ref{fig:singVals_critLayer}(e), the secondary flow decreases the amplification levels in comparison to the streamwise turbulent case. Additionally, we again see that the secondary flow has a notable effect on the forcing modes, moving them closer to the corners of the domain. 
 Additionally, the different critical layer locations between the laminar and turbulent systems, seem to be the reason for the discrepancies between the appearance and distribution of the modes.

\begin{figure}[ht!]
\centering {
\vspace{-0.3cm}
{\hspace*{-2.4cm}\includegraphics[width= 1.3\textwidth]{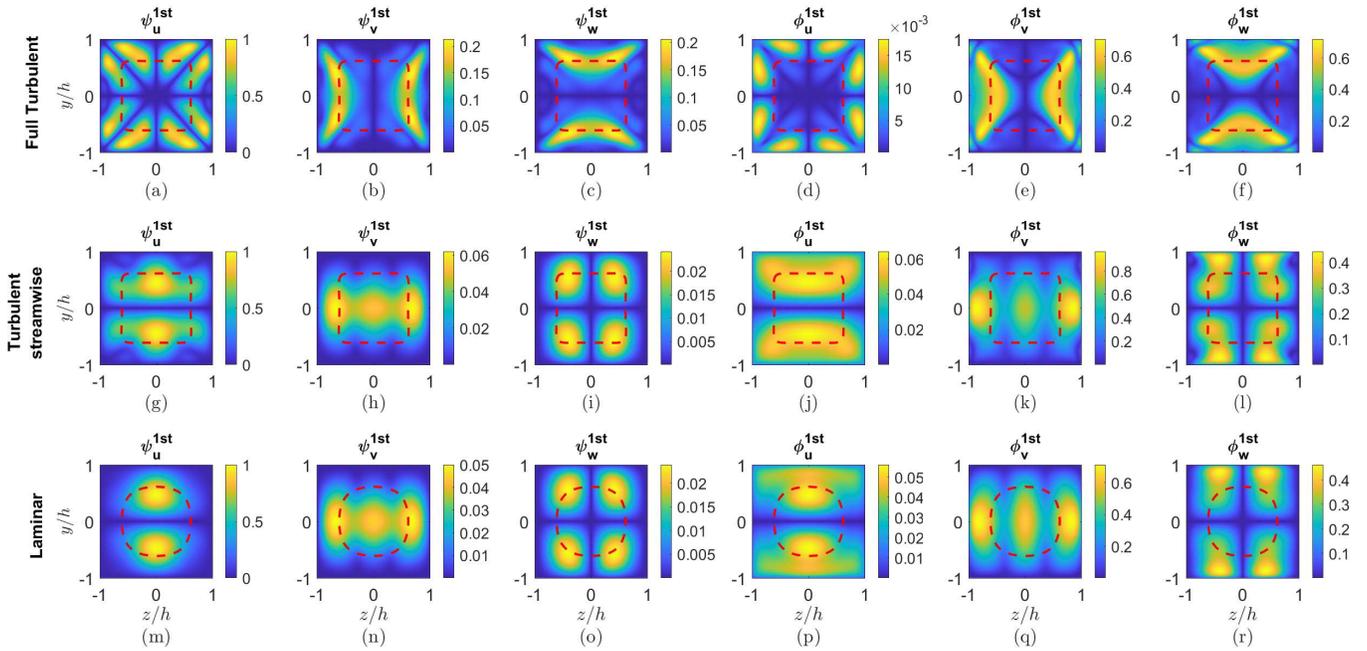}} }
\vspace{-0.4cm}
\caption{Normalized amplitude of the optimal response $\psi(y,z)$ and forcing $\phi(y,z)$ modes of the resolvent operator for a laminar pressure-driven (a)-(f), full turbulent (g)-(l) and streamwise component of a turbulent (m)-(r), square duct flow with with $Re_b=2500$, $k_x=0.1$ and $c=1.15$. Critical-layer locations indicated by red dashed lines.
}
\label{fig:fullModes_All_kx0p1c1p15}
\end{figure}

Fig.~\ref{fig:fullModes_All_kx0p1c1p15} depicts a configuration in which both the singular values and the resolvent modes of the laminar and streamwise turbulent systems are extremely similar, while the full turbulent system has an entirely distinct mode structure, with approximately double the amplification (per Fig.~\ref{fig:singVals_critLayer}(h)). In particular, rather than having a large streamwise velocity response in near the $z = 0$ plane, the streamwise velocity response for the full turbulent mean case is concentrated near the corners. The forcing modes for this case consist of $v$ and $w$ components that appear to generate motions approximately parallel to the sidewalls. While not shown, the suboptimal modes for the full turbulent case resemble the leading modes for the other cases.

\begin{figure}[ht!]
\centering {
\vspace{-0.5cm}
{\hspace*{-2.4cm}\includegraphics[width= 1.3\textwidth]{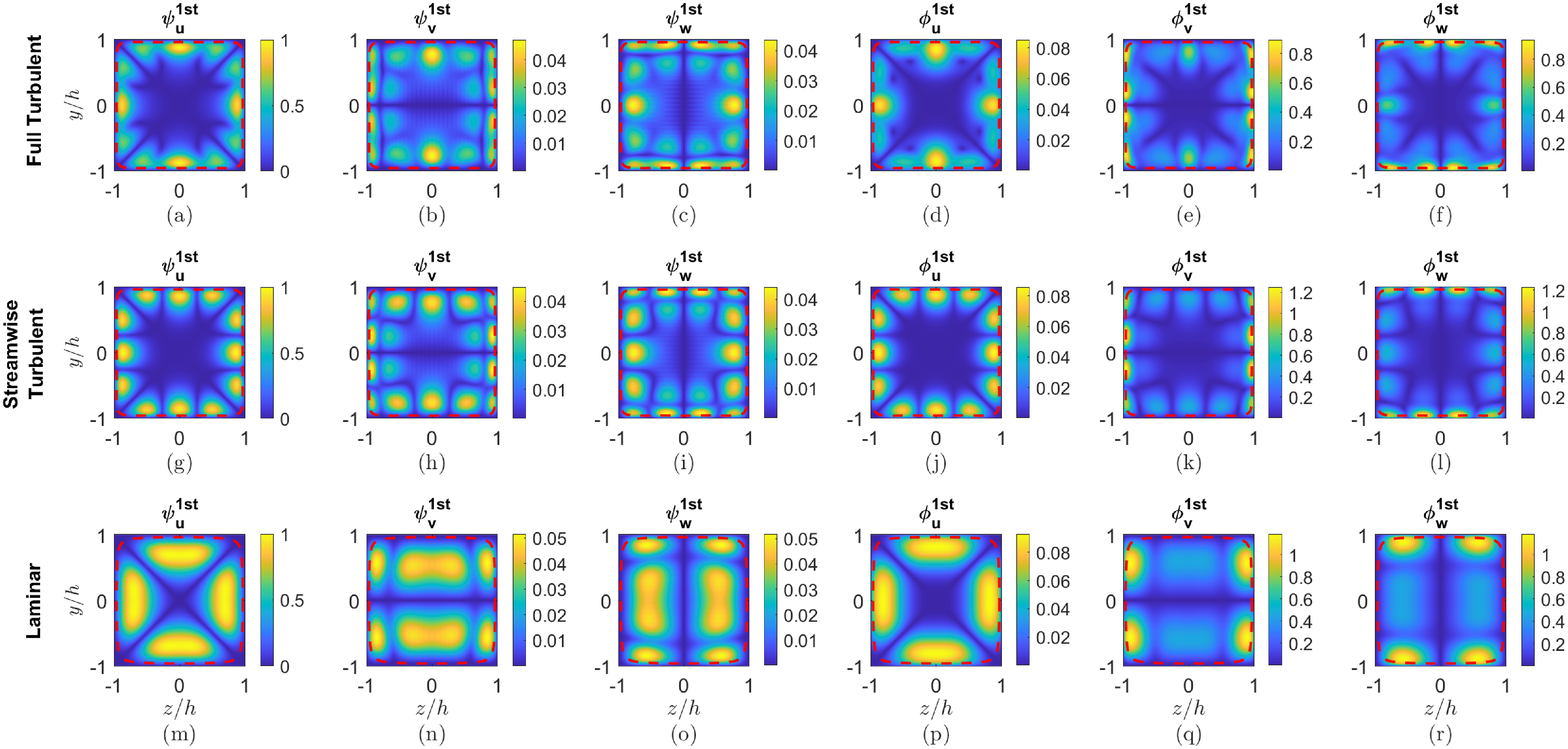}} }
\vspace{-0.8cm}
\caption{Normalized amplitude of the optimal response $\psi(y,z)$ and forcing $\phi(y,z)$ modes of the resolvent operator for a laminar pressure-driven (a)-(f), full turbulent (g)-(l) and streamwise component of a turbulent (m)-(r), square duct flow with with $Re_b=2500$, $k_x=0.1$ and $c=0.45$. Critical-layer locations indicated by red dashed lines.
}
\label{fig:fullModes_All_kx0p1c0p45}
\end{figure}

Fig.~\ref{fig:fullModes_All_kx0p1c0p45} shows a case where the critical layer is close to the wall. In this case, the laminar case gives a distinctly different mode shape, likely due to the vastly different mean velocity profile in the near-wall region. The two turbulent cases share similar mode structures, though with some subtle differences. Both response modes are again dominated by the streamwise component, with the forcing dominated by  wall-parallel $v$ and $w$ velocity components. For the streamwise-only turbulent mean, the streamwise response mode consists of three lobes on each wall, of approximately equal strength. For the case with secondary mean flow, the center lobe on each wall is substantially larger in magnitude. From Fig.~\ref{fig:singVals_critLayer}(a), this yields larger amplification than the streamwise-only case. Note that for turbulent duct flows, the turbulence intensity is largest in the near-wall regions in the center of each side.

\begin{figure}[ht!]
\centering {
\vspace{-0.35cm}
{\hspace*{-2.4cm}\includegraphics[width= 1.3\textwidth]{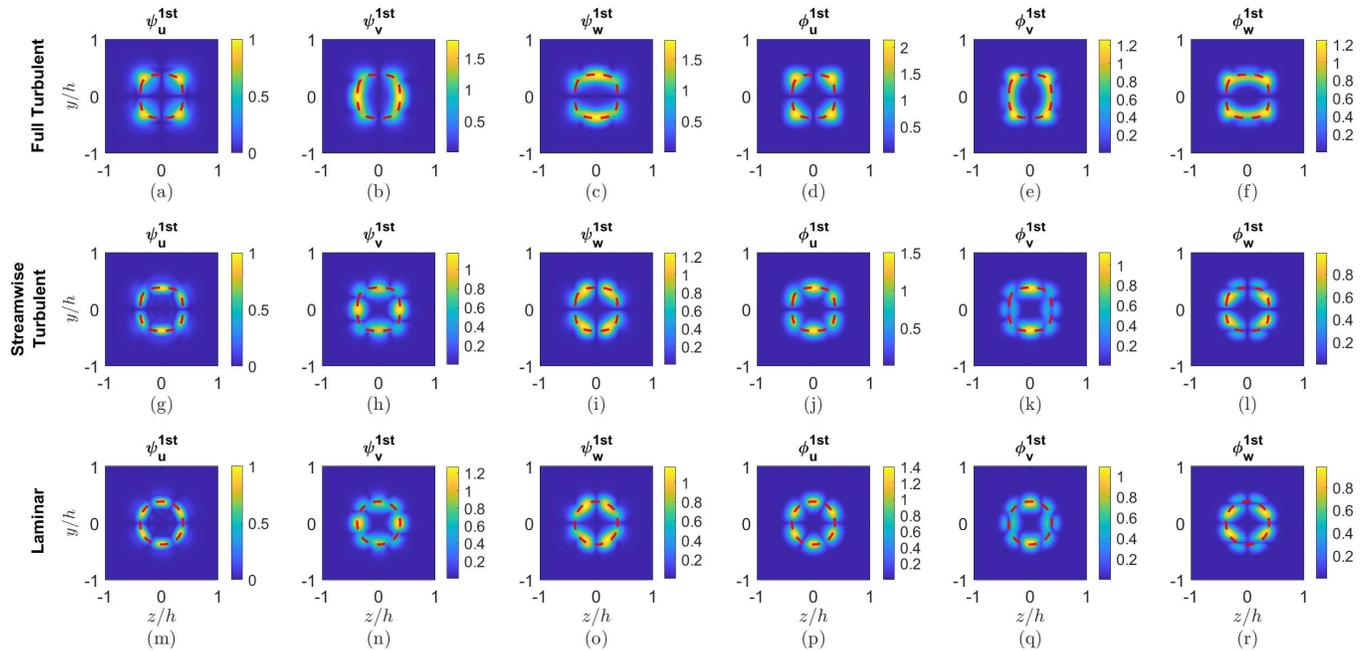}} }
\vspace{-0.8cm}
\caption{Normalized amplitude of the optimal response $\psi(y,z)$ and forcing $\phi(y,z)$ modes of the resolvent operator for a laminar pressure-driven (a)-(f), full turbulent (g)-(l) and streamwise component of a turbulent (m)-(r), square duct flow with with $Re_b=2500$, $k_x=10$ and $c=1.25$. Critical-layer locations indicated by red dashed lines.
}
\label{fig:fullModes_All_kx10c1p25}
\end{figure}

Fig.~\ref{fig:fullModes_All_kx10c1p25} considers modes which, with the selected wavenumber $k_x=10$, produces structures that are very localized about a critical layer that is close to a circle for both laminar and turbulent systems. Despite the fact that these modes are far away from regions of large secondary flow, there is still a noticeable difference in the mode shape and amplification.



\section{Discussion and Conclusions}
\label{sec:disc}
This work has focused on studying linear energy-amplification mechanisms that arise in incompressible flows through square and rectangular ducts. Particular attention has been given to isolating the effects that secondary flows have on such mechanisms. This was achieved by considering a turbulent mean velocity profile where the flow in the non-streamwise directions was set to zero. While such a configuration is not physical, it allows for the direct effects of the secondary flows to be isolated. Interestingly, it was found that the presence of secondary flows tended to increase amplification for structures with very large ($k_x = 0.1$) and very short ($k_x = 10$) streamwise extent, but that secondary flows reduced amplification for intermediate ($k_x = 1$). Secondary flows were found to change the structure of the leading resolvent forcing modes more so than the response modes, often giving forcing modes that were closer to the corners of the domain, and further away from where the response modes were concentrated. We additionally identified one case ($c = 1.15$, $k_x=1$) where an entirely distinct mode emerged in the presence of secondary flow, with approximately double the amplification. The fact that these findings emerged with secondary flows on the order of 1\% of the primary streamwise flow suggests that it could be important to consider their effect even in  configurations where the mean flow is largely unidirectional. 
Note that the sensitivity of linear modes to distortion in the base flow has also been considered in the context of (spanwise homogeneous) high-speed boundary layers \cite{park2019sensitivity}.


Further research directions include performing a more comprehensive 
sweep to map out the regions in parameter space where secondary flow enhances and suppresses linear amplification, extending the analysis to higher Reynolds numbers, and comparing the modes identified to structures established from time-resolvent data from turbulent duct flow simulations. It could additionally be interesting to studying the effect of secondary flows in pipes with curvature (Prandtl's secondary flow of the first kind), which can lead to more extreme transient events \cite{chin2020backflow}. 

\section*{Acknowledgments}
STMD acknowledges support from the Air Force Office of Scientific Research grant FA9550-22-1-0109. RV acknowledges financial support by the Swedish Research Council (VR).



\bibliography{Master}

\end{document}